\documentclass{aa}  

\usepackage{graphicx}
\usepackage{txfonts}
\newcommand{\MH}{\mathrm{[M/H]}} 
\newcommand{\A}{A_0} 
\newcommand{\D}{D} 
\newcommand{\Os}{O_j} 
\newcommand{\Pe}{\hat{P}} 
\newcommand{\Dmax}{D_\mathrm{max}} 
\newcommand{\Dmin}{D_\mathrm{min}} 
\newcommand{\Shrd}{S'} 
\newcommand{\bpmrp}{G_{BP}-G_{RP}}
\newcommand{\BP}{G_{BP}}
\newcommand{\RP}{G_{RP}}

\newcommand{\hrdpoint}{i}

\usepackage{color}

\begin{document}

   \title{FEDReD I: 3D extinction and stellar maps by Bayesian deconvolution}

   \author{C. Babusiaux \inst{1,2}         
          \and
          C. Fourtune-Ravard \inst{2}
          \and 
          C. Hottier \inst{2} 
          \and 
          F. Arenou \inst{2}
          \and
          A. G\'omez \inst{2}
          }

  \institute{Univ. Grenoble Alpes, CNRS, IPAG, 38000 Grenoble, France
              \email{Carine.Babusiaux@univ-grenoble-alpes.fr}
              \and 
	    GEPI, Observatoire de Paris, CNRS, Universit\'e Paris Diderot ; 5 Place Jules Janssen 92190 Meudon, France
             }

   \date{Received ; accepted }

  \abstract
   {While Gaia enables to probe in great detail the extended local neighbourhood, the thin disk structure at larger distances remains sparsely explored. } 
   {We aim here to build a non-parametric 3D model of the thin disc structures handling both the extinction and the stellar density simultaneously.}
   {We developed a Bayesian deconvolution method in two dimensions: extinction and distance. It uses a reference catalogue which completeness information defines the selection function. It is designed so that any complementary information from other catalogues can be added. It has also been designed to be robust to outliers, frequent in crowded fields, and differential extinction. The prior information is designed to be minimal: only a reference H-R diagram. We derived for this an empirical H-R diagram of the thin disk using Gaia DR2 data and synthetic isochrone-based H-R diagrams can also be used. }
   {We validated the method on simulations and real fields using 2MASS and UKIDSS data complemented by Gaia DR2 photometry and parallaxes. We detail the results of two test fields: a 2MASS field centred around the NGC~4815 open cluster which shows an over-density of both extinction and stellar density at the cluster distance, and a UKIDSS field at $l=10\degr$ where we recover the position of the Galactic bar.}
   {} 
  
   \keywords{dust, extinction --
		Galaxy: structure --
		ISM: structure --
                Methods: statistical 
               }

   \maketitle
%

\section{Introduction}

Uncovering the Galactic structure within the Galactic plane is a challenging issue due to the mix between stars and dust at different distances, the dust affecting the light of the stars through the extinction. 

Several methods have now been developed to draw 3D extinction maps. Fully model based \citep[e.g.][]{Drimmel01}, using a stellar distribution model but deriving a non-parametric 3D extinction map \citep[using the Besan\c{c}on model]{Marshall06,Chen13,Schultheis14}, the distribution of stars near the main-sequence turn-off \citep{Gontcharov17} or, the most common, using individual stellar distance and extinction estimates which are then inverted \citep{Arenou92, Vergely10, Lallement19, Berry12, Chen14, Hanson14, Rezaei17, Chen19, Anders19}. 
\cite{Green14} samples the full probability density function of distance, reddening for individual stars, derived on main-sequence star's broad band photometry, to build their 3D extinction map, taking into account the survey selection function. 
\cite{Sale12} uses a full hierarchical model to handle simultaneously the mean-distance-extinction relationship for a sightline and the individual stellar properties. 

To derive stellar density distributions, most methods are parametric \cite[e.g.][]{Drimmel01,Reyle09}. Non-parametric stellar density models have been derived up to now when the extinction could be handled independently, e.g. assuming that most of the extinction occurs in the foreground of the structure under-study: at high galactic latitudes \citep[e.g.][]{deJong08} and for the bulge structure outside of the Galactic plane \citep[e.g.][both using deconvolution methods]{LopezCorredoira00,Wegg13}. Other methods specifically studying the bar structure have been searching for the red clump position using a magnitude independent of extinction \citep[e.g.][]{Stanek94,Babusiaux05,Nishiyama05,CabreraLavers08,Wegg15}.

Here we wish to work within the Galactic plane and derive the non-parametric distribution of both the extinction and the stellar density at the same time. This is the first time this is attempted in the Galactic disk. For this, we use a Bayesian deconvolution method \citep{Richardson72, Lucy74} using all the stellar information available within a given line of sight. We present here the algorithm we developed, FEDReD (Field Extinction - Distribution Relation Deconvolver). It is designed to work using one reference catalogue on which the completeness model will be based and any other survey which can provide complementary information on the stars observed. We choose in the description and applications presented here to use near-infrared surveys such as 2MASS \citep{Skrutskie06} or the UKIDSS Galactic Plane Survey \citep{Lucas08} as reference catalogues as they can probe large distances in high extinction fields and Gaia data \citep{Prusti16,GDR2paper} as complementary information. 
We present in Sect.~\ref{sec:method} the method, in Sect.~\ref{Shrd} the H-R diagram (HRD) priors we constructed and in Sect.~\ref{sec:tests} results in both a simulated field and selected test fields.


\section{\label{sec:method}Method}

\subsection{\label{Sdeconv}Bayesian deconvolution}

We wish to derive the probability distribution $P(\A,D)$ which gives the probability of a star to have both an interstellar extinction $\A$ \citep[extinction at 550~nm, which is roughly the centre of the $V$ band, e.g.][]{CBJ11} and a distance $D$ along a given line of sight.
$P(\A|D)$ gives the variation of the extinction with distance and $P(D)$ gives the stellar density distribution along the line of sight. 

We first assume that we are  observing all the N stars along a given line of sight, each star observed $\Os$ having several observables (here as a minimum NIR magnitudes and potentially optical magnitudes and parallax). 
What we which to derive is

\begin{equation}
P(\A,D) = \sum_{j=1}^{N} P(\A,D|\Os) P(\Os) 
\end{equation}

The sum is discrete instead of the usual integral as we are observing a finite number of stars. 

We have through Bayes' theorem: 
\begin{equation}
\begin{aligned}
 P(\A,D|\Os) &\quad=  \frac{P(\Os|\A,D)\,P(\A,D)}{P(\Os)} \\
 &\quad= \frac{P(\Os|\A,D)\,P(\A,D)}{\int_{(\A,D)} P(\Os|\A,D)\,P(\A,D)\,d\A\,dD}
 \end{aligned}
 \label{eq:PADO}
\end{equation}

Following the well-known Richardson-Lucy deconvolution algorithm, we can estimate $P(\A,D)$ by iteratively computing $h(\A,D)$ \citep[][with $\xi=A,D$ and $x_n=O_j$]{Lucy74}:
\begin{equation}
h_{k+1}(\A,D) = \frac{1}{N} \sum_j \frac{P(\Os|\A,D)\,h_{k}(\A,D)}{\int_{(\A,D)} P(\Os|\A,D)\,h_{k}(\A,D)\,d\A\,dD} 
\label{eq:deconv}
\end{equation}
The initial values $h_0(\A,D)$ are discussed in Sect.~\ref{Sinit}.

However we do not observe all the stars, but we can model (Sect. \ref{Scompl}) the selection function ($S$) through a model of the completeness of our near-infrared data: $P(S|m_J,m_H,m_K)$. 
What we can compute iteratively is then in fact $P(\A,D|S)$: 

\begin{equation}
P(\A,D|S)= \sum_{j} P(\A,D|\Os,S)\,P(\Os|S) 
\end{equation}
still with $\Os$ being an observed star with at minimum NIR magnitudes observables. $P(\Os|S)$ is the probability of an observed star to be in the selected sample. 
Similarly to Eq.~\ref{eq:PADO} we have
\begin{equation}
 P(\A,D|\Os,S) =  \frac{P(\Os|\A,D,S)\,P(\A,D|S)}{\int_{(\A,D)} P(\Os|\A,D,S)\,P(\A,D|S)\,d\A\,dD}
\end{equation}
with
\begin{equation}
 P(\Os|\A,D,S) = \frac{P(S|\Os,\A,D)\,P(\Os|\A,D)}{P(S|\A,D)}
\end{equation}
The observed star being actually observed, $P(S|\Os)=1$.
We therefore estimate $P(\A,D|S)$ by iteratively computing $h(\A,D|S)$:
\begin{equation}
h_{k+1}(\A,D|S) = \frac{1}{N} \sum_{j} \frac{\psi_{k}(\A,D)}{\int_{\A,D} \psi_{k}(\A,D)\,d\A\,dD} 
\end{equation}
with $\psi_{k}(\A,D) = P(\Os|\A,D)\,h_{k}(\A,D|S) / P(S|\A,D)$, all the observed stars contributing with the same weight to the selected sample.

At the last iteration $K$ (see Sect.~\ref{Sconv} for the convergence criteria), we have an estimate of $P(\A,D|S)$ which we will note in the following $\Pe(\A,D|S) = h_K(\A,D)$.
We can then retrieve an estimate of $P(\A,D)$ with
\begin{equation}
 \Pe(\A,D) \propto \frac{\Pe(\A,D|S)}{P(S|\A,D)}
\end{equation}

We assume here that all sources in the catalogue are real stars, which is unfortunately not the case, in particular in crowded fields where false detections and false cross-match between observations in different filters and catalogues can be numerous. Those false observations imply to use a robust method to derive $\Pe(\A|D)$ while they should not impact significantly $\Pe(D)$.

\subsection{Individual probabilities $P(\Os|\A,\D)$}

To derive each star's probability $P(\Os|\A,\D)$, we compare its observables to the properties of all points of an intrinsic HRD, either isochrone-based or empirical (see Sect.\ref{ShrdIsoc} for details on the HRD) .

A given HRD point $i$ with an absolute magnitude $M_i$ at a distance $\D$ and with an extinction $\A$, has an apparent magnitude $m_i$ :
\begin{equation}
m_i = M_i + 5 \log \D -5 + k_m(i,\A) \A.
\label{eq:Pogson}
\end{equation}
where $k_m$ is the extinction coefficient in the given $m$ photometric band.
We take into account the fact that $k_m$ is actually a function of the intrinsic colour of the star (known through $i$) and of the extinction itself $\A$ through the formalism of \cite{Danielski18}, using the same coefficients as \cite{Lallement19}:
\begin{equation}
k_m(i,\A) = a_1 + a_2 X_i + a_3 X_i^2 + a_4 X_i^3 + a_5 \A + a_6 \A^2 + a_7 X_i \A
\label{eq:extcoef}
\end{equation}
where $X_i$ is by default the $G-K$ colour of the HRD point $i$. If the HRD is based on isochrones, $X_i$ can be chosen to be the stellar temperature. 

\begin{equation}
\begin{aligned}
P(\Os|\A,\D) &\quad= \sum_i P(\Os|\A,\D,i)\,P(i) \\
 &\quad= \sum_i \prod_{m\in\{J,H,K\}} P(\tilde{m}|m_i)\,P(\tilde{\varpi}|\varpi)\,P(i)
\end{aligned}
\end{equation}

To compute $P(\tilde{m}|m_i)$ and $P(\tilde{\varpi}|\varpi)$ we assume Gaussian observational errors on the magnitudes for the NIR surveys, on the flux for Gaia, and on the parallax. 

We derive $P(\Os|\A,\D)$ for a thin 2-D grid of distances and extinction. The distance being computed using the magnitudes, we do not use a constant step in distance but in distance modulus $\mu$ with a step of 0.05 mag, corresponding to the typical photometric error of the input catalogues. We therefore work in $d\mu = 5/\log(10)\,dD / D$. Similarly, we choose a step in extinction $\A$ of 0.05~mag. The grid typically extends from 0.1 to 30~kpc in distance and from 0 to 30~mag in extinction, although this can be adapted to the field of view and the survey to optimise the computation time. 
Illustrations of the results for different stellar types are provided in Fig.~\ref{fig:indivPAD}. It shows that the information is mostly carried by red clump stars and that the Gaia parallax and/or photometry is needed to differentiate a red clump star from a red dwarf.

\subsection{\label{Scompl}Selection function model $P(S|\A,D)$}

As previously, we compute the probability to be selected using the H-R diagram prior:
\begin{equation}
 P(S | \A,D) = \sum_i P(S | \A,D,i)\,P(i) = \sum_i \prod_{m \in \{J,H,K\}} P(S|m_i)\,P(i)
\end{equation}

We adopted the following model for the completeness of the surveys which is generic enough to reproduce simply typical completeness curves:
\begin{equation}
\label{eq:compl}
P(S|m) \propto  \left \{
  \begin{array}{l l}
1 - \exp\left[-\left(\frac{m^{*} - m}{\beta}\right)^\alpha\right] & \mbox{if } m \leq m^{*} \\
0 & \mbox{if } m > m^{*}
  \end{array}
  \right .
\end{equation}
$P(S|m)$ is the probability for a star to be observed in a given photometric band given its magnitude true $m$. $\alpha$, $\beta$ and $m^{*}$ are three parameters defining the completeness function which depends on the survey and on the crowding of the field of view.
We used simulations to derive the parameters of this completeness function for the different surveys. We simulated a few typical fields with the \cite{Marshall06} extinction model, the \cite{Fux99} stellar density distribution and modelling the errors from the observed catalogues. We fitted the parameters $\alpha$, $\beta$ and $m^{*}$ of Eq.\ref{eq:compl} using the observed $J$, $H$ and $K$ magnitude distributions $P(m|S)$ and the simulated ones $P(m)$  and solving $P(S|m) \propto P(m|S) / P(m)$ on their cumulative distribution functions.
We found that $\alpha=2$ and $\beta=2$ were globally appropriate for the UKIDSS survey and a sharper curve with $\alpha=10$, $\beta=1$ for the 2MASS survey, in agreement with the studies of \cite{Lucas08} for UKIDSS and \cite{Skrutskie06} for 2MASS.   
We found that for both surveys, a reasonable approximation of $m^{*}$ can be obtained by adding 2~magnitudes to the maximum of the observed magnitude distribution $P(m|S)$. 
We note that this approximation is only valid when the distribution of stars is relatively smooth. For example, it is not valid anymore when a large stellar density is present near the end of the completeness survey, e.g. typically in fields dominated by the bar feature. In those fields parameters must be adjusted either through simulations, or better, through direct image completeness tests (e.g. \cite{Surot19}).  

For the UKIDSS data, photometric errors can go very high so we restricted the data used to stars with photometric errors lower than 0.1~mag. This means in practice restricting UKIDSS photometry to be roughly within $J<19$, $H<18$, $K<17$. We take this truncation into account in our selection function model. For this, we first model the photometric errors in the band $m$ by a fit on the observables.
\begin{equation}
 \sigma(m) = a + b e^{c m}
 \label{eq:sigm}
\end{equation}
As the errors in the UKIDSS survey are a direct function of the observed magnitude, we derive from this the magnitude $m_\sigma$ corresponding to our truncation on $\sigma$. We then have the probability of a theoretical star of magnitude $m$ to be selected through the cumulative distribution function of the magnitude errors $\sigma_m$ derived with Eq.~\ref{eq:sigm}: 
\begin{equation}
 P(S|m) = P(\tilde{m} < m_\sigma | m, \sigma(m))
 \label{eq:sigcut}
\end{equation}
For UKIDSS the global selection probability is then the product of Eq.~\ref{eq:compl} and \ref{eq:sigcut}.

\subsection{\label{Sinit}Initial values}

The construction of the initial value of the iteration, $h_0(\A,D|S)$, needs two initial conditions: $P_0(D)$ and $P_0(\A|D)$. Then it is simply computed as:
\begin{equation}
 h_0(\A,D|S) \propto P(S|A,D) P_0(\A|D) P_0(D) 
\end{equation}

\subsubsection{$P_0(D)$}

We only take into account the cone effect on a constant underlying stellar density profile: $P_0(D) \propto D^2 $. 
We tested also using a disc exponential profile start but that does not influence at all our results and we therefore stay with the simple flat start. 

\subsubsection{$P_0(\A|D)$}

Here also we use a flat start : $P_0(\A|D)=1$. 

However, a number of degenerate solutions occur due to the confusion between giants and red dwarfs (see Fig.~\ref{fig:indivPAD}), especially in crowded fields with high extinctions where there are not enough Gaia DR2 stars to constrain the solution at small distance (typically below 1~kpc but depending on the cone aperture chosen). To avoid this, we tested that adding an extra simple local prior could help, e.g. that the extinction cannot be too high locally: $P(\A|D)=0$ for $\A>10 D$, with $D$ the distance in kpc. The local map of \cite{Lallement19} confirms that this is a very safe prior. But the algorithm is robust enough so that it is not needed in practice, at least in well behaved fields. For UKIDSS fields, where the red clump information starts only at relatively large distances and with very few Gaia information, adding this simple local prior is sometimes not enough and using a prior based on 2MASS data over a larger field of view is needed. However such a prior should be robust enough to differential extinction in order to be used safely. 

\subsection{\label{Sconv}Convergence}

Assessing the convergence of such a deconvolution is always tricky. We decided to stop the iterations when the convergence rate slows down. 
\begin{equation}
\Delta_k = \sum_{\A,D} [ h_k(\A,D|S) - h_{k-1}(\A,D|S) ]^2
\end{equation}
\begin{equation}
cr = \vert \frac{\Delta_{k+1}-\Delta_{k}}{\Delta_{k}} \vert < 0.05
\end{equation}
The 0.05 threshold for the convergence criteria is somewhat arbitrary but was tested on both simulations and real data to enable to reach the final shape of the $\Pe(\A|D)$ relation without introducing too much noise in the overall $\Pe(\A,D)$.
We limited the number of iterations to 50, which was reached in a few crowded areas only. We checked on those areas that, although the resulting matrix $\Pe(\A,D)$ is quite noisy, the algorithm to recover $\Pe(\A|D)$ (Section \ref{SgetAD}) is robust enough to cope with these areas. 

\subsection{\label{SgetAD}Deriving $\A(D)$}

\begin{figure}[t]
\centering
\includegraphics[width=\columnwidth]{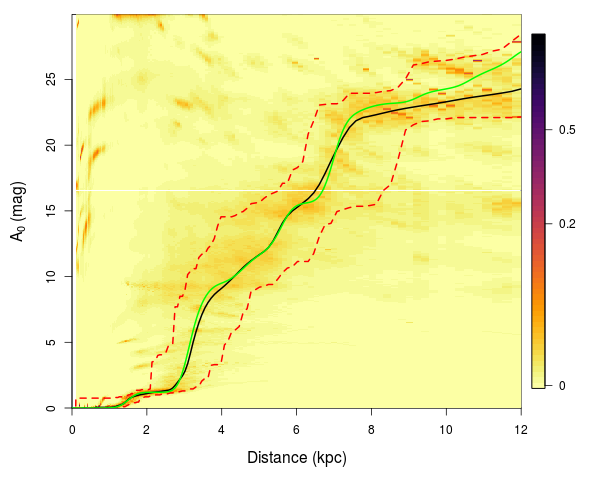}
\caption{$\Pe(\A|D)$ from the first deconvolution. The green line is the real relation of the simulation. The black line is the first deconvolution resulting $A(D)$ estimated relation (constrained median cubic spline fit on the MCSs). The red dotted lines are the minimum/maximum envelop of the MCSs. For the second deconvolution, $P_0(\A,D)$ is initialised with a zero probability outside of this red envelop.}
\label{fig:resmat1outliers}
\end{figure}

We derive $\Pe(\A|D)$ from $\Pe(\A,D)$ obtained at the end of the deconvolution process with
\begin{equation}
\Pe(\A|D) \propto \frac{\Pe(\A,D)}{\Pe(D)} = \frac{\Pe(\A,D)}{\int_{\A} \Pe(\A,D)\,d\A}  
\end{equation}
We now search for the relation $\A(D)$ corresponding to the maximum of the probability $\Pe(\A|D)$ with the physical constraint that $\A(D)$ should increase with $D$. 
For this, we randomly generate 10\,000 monte-carlo solutions of $\A(D)$ (called MCS hereafter) according to the following algorithm. 
We randomly select a distance $D_l$ within our working area following the probability weights $\Pe(D|S) = \int_{\A} \Pe\A,D|S)\,d\A$. 
For each distance $D_l$ we randomly select a corresponding extinction within the possible values allowed by the increasing constrain and the extinctions already assigned to the previous distances. This random selection of $\A(D_l)$ is done using the probability weights defined by $\Pe(\A|D_l)$. We initiate the generation by setting $\A(0)=0$ and $\A(\Dmax)=A_{max}$. 
We compute the total log-likelihood of a MCS by $\log(\mathcal{L})=\sum_l \log(\Pe(\A(D_l),D_l))$.
We then select the best 1\,000 MCSs. We re-generate 10\,000 random MCSs but this time further constraining the solutions to be within the extinction envelop of the 1\,000 best MCSs for each distance. If the log likelihood of the new generated solution is better than the worse of the MCSs, the new solution replaces it. 
Finally a median cubic spline fit with an increasing constrain \citep{cobs} is applied on the final 1\,000 MCSs. Those solutions are illustrated in Fig.~\ref{fig:resmat1outliers} for our default simulation (Sect.~\ref{Ssimu}).
The 68\% confidence interval (equivalent to 1-$\sigma$ for a normal distribution) is derived from the quantiles of the MCSs distributions. 

As illustrated in appendix~\ref{Sannex}, red clump stars are providing the strongest constraints on the distance/extinction distribution thanks to their small intrinsic dispersion in absolute magnitude and colour. To avoid noise induced by degenerate solutions, we therefore apply the previously described procedure to select the best MCSs on the distance interval where red clump stars are expected to provide information. To do so, we used the red clump absolute magnitudes of \cite{RuizDern18} and considered when the red clump is expected to saturate to set the minimum distance and to reach the completeness limit to set the maximum distance. For the first generation of MCSs, the log-likelihoods are computed only within the red clump distance range derived assuming no extinction. For the second generation the distance range is updated using the maximum and minimum extinctions of the best MCS. 
The resulting MCS results are considered valid within the final distance interval $\Dmin$/$\Dmax$ still defined by the expected red clump saturation and completeness limit, using this time the extinction provided by the 1-$\sigma$ confidence interval of the derived $\Pe(\A|D)$. 

A second deconvolution process is done using the first one as the initial probability distribution, e.g. setting $P_0(\A,D)$ with a Gaussian distribution according to the MCSs quantiles, zero outside the MCSs envelop, and a flat probability outside $\Dmin$/$\Dmax$ just taking into account the maximum / minimum (respectively) extinction of the last valid distance. This new starting distribution removes in particular the degenerate solutions seen at small distances with large extinction due to the confusion between giants and red dwarfs. Such a second step is needed mainly to derive accurately the distance distribution, the first pass deriving already well the extinction/distance relation. 
A last restriction on the definition of our estimated distance validity range for our results is the addition of a constraint on $\Dmin$/$\Dmax$ to exclude distances too far away or too close within our cone angle to have enough stars: $\Pe(D<\Dmin~|~S)>0.01$ and $\Pe(D<\Dmax~|~S)<0.99$. This last restriction is needed in particular for anticentre areas with a low intrinsic stellar density at large distance. 

\begin{figure}[t]
\centering
\includegraphics[width=\columnwidth]{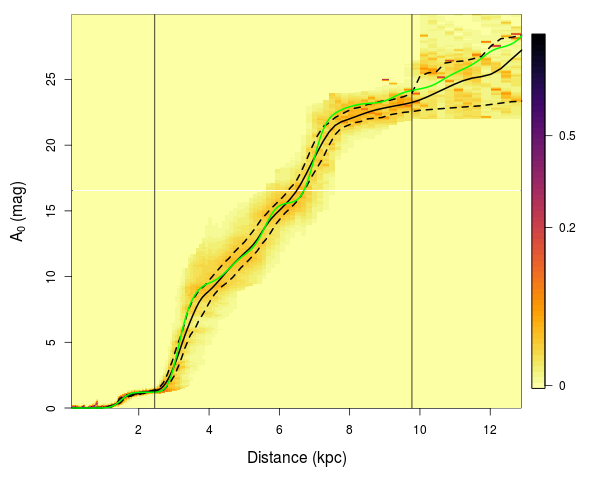}
\caption{$\Pe(\A|D)$ from the second deconvolution. The black line is the final $A(D)$ estimated relation. Dotted lines corresponds to the 1-$\sigma$ CI. The green line is the real relation of the simulation. The vertical lines correspond to $\Dmin$ and $\Dmax$, e.g. show the valid distance range derived from the red clump star saturation magnitude and completeness limit respectively.}
\label{fig:resmat2final}
\end{figure}

\subsection{\label{SgetProbaD}Deriving $\Pe(D)$}

\begin{figure}[t]
\centering
\includegraphics[width=8cm]{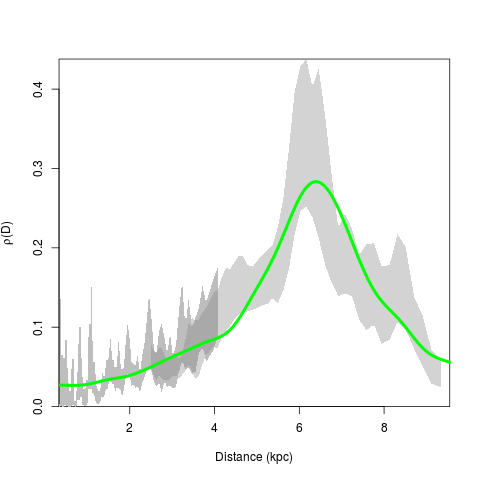}
\caption{1-$\sigma$ confidence interval of the derived stellar density $\rho(D)$ as obtained by
  bootstrap obtained from two mock catalogues of the same area at l=10$\degr$, dark
  grey: 2MASS like,  light grey: UKIDSS like. 
 Green: input simulation stellar density.
}
\label{fig:rhoSimu}
\end{figure}

We simply compute $\Pe(D) = \int_{\A} \Pe(\A,D)\,d\A$. However the result is quite noisy, as usual for Richardson-Lucy deconvolution,  and recovering its error bar is not obvious. We chose here to estimate the confidence interval using a simple bootstrap method on the second deconvolution. For this, we bootstrap the input stars and the first deconvolution prior. For the latter, we bootstrap the MCSs, we select a random one and put as prior a flat distribution within the 2-$\sigma$ interval defined by the MCS centred around this random MCS. 
We then remove the cone effect to recover $\rho(D) \propto \Pe(D) / D^2$. The result can be seen in Fig.~\ref{fig:rhoSimu} for 2 mock catalogues of the same field with 2MASS and UKIDSS photometric properties (see Sect.~\ref{Ssimu}).  
We only fit $\rho(D)$ within the valid distance range as defined previously, e.g. where red clump stars are within the rough completeness regime in all three NIR bands. 
The original $P(D)$ is recovered in all cases within 2 sigma but quite noisy, in particular at small distances. 

\section{\label{Shrd}H-R diagram priors}

We implemented two different H-R diagram priors, an empirical one based on Gaia observations and a theoretical one based on isochrones.
One advantage of the Gaia empirical HRD is that it does not need IMF, metallicity nor age priors and takes into account naturally the presence of binaries. However the theoretical HRD based on isochrones is still useful if we want to add other constraints than parallax and photometry, e.g. spectroscopic information, or if we want to test the impact of variations of the HRD within the Galaxy. An other motivation for implementing an empirical HRD is the known mismatch between the atmosphere models used in the isochrones and the observed intrinsic colour-colour relations, in particular for cool stars \citep[e.g.][]{Aringer16, RuizDern18}.

\subsection{\label{ShrdGaia}The Gaia Empirical HRD}

We use by default an empirical HRD based on Gaia DR2. We restrict ourselves to a distance above the plane $\vert Z \vert < 50$~pc as we are looking here only in the galactic plane. This value is a trade-off between having enough stars to sample the giant branch and staying as close as possible to the Galactic plane, e.g. within a relatively homogeneous stellar population mixture. 
We select only low extinction stars to build the empirical HRD to avoid adding uncertainties associated with any local extinction map, but as a consequence this HRD can only be used in regions with relatively high extinction so that our extinction residuals become negligible, which is the case for the Galactic disk fields for which this HRD has been built.
We applied the same astrometric and photometry filters as in Appendix B of \cite{DPACP-31} with the exception of the photometric flux error one 
\footnote{
   parallax\_over\_error $>$ 10 \\
   phot\_bp\_rp\_excess\_factor $<1.3+0.06 $ bp\_rp$^2$ \\
   phot\_bp\_rp\_excess\_factor $> 1.0+0.015$ bp\_rp$^2$ \\
   visibility\_periods\_used$>8$ \\
   astrometric\_chi2\_al/(astrometric\_n\_good\_obs\_al-5)$<$1.44 max(1,exp(-0.4*(phot\_g\_mean\_mag-19.5)))
}.

Instead we used a sharp limit in magnitude: $G<20$, $\RP<19$ and $\BP<18$ to have a simple selection function model. To select low extinction stars we also used the 3D extinction map of \cite{Capitanio17}, but to get enough red giants close to the plane we used the rather large limit of $E(B-V)<0.05$~mag. 
We use this sample to build an Hess Diagram on a grid of $\bpmrp$ colour (step 0.01~mag) and $M_G$ (step 0.02~mag), e.g. $P(M_G,\bpmrp|\Shrd)$, with $\Shrd$ the HRD stars selection function. 
To correct for the selection function we derive 
\begin{equation}
P(M_G,\bpmrp) \propto \frac{P(M_G,\bpmrp | \Shrd)}{P(\Shrd | M_G,\bpmrp)} 
\end{equation}
As previously, we call $\hrdpoint$ one of those HRD point ($M_G,\bpmrp$).

\begin{equation}
 P(\Shrd | \hrdpoint) = 
 \iiint_{d,l,b} P(\Shrd | \hrdpoint,d,l,b) P(d,l,b) \,dd \,dl \,db
\end{equation}

We assume an homogeneous sky distribution, which is a good enough approximation for our work within the $\vert Z \vert < 50$~pc constraint of our local sample. We therefore have $P(d,l,b)=d^2 \cos(b)$. 
We move our integral in parallax space instead of distance as this is our observable:

\begin{equation}
 P(\Shrd | \hrdpoint) = \iiint_{\varpi,l,b} P(\Shrd | \hrdpoint,\varpi,l,b) \cos(b) / \varpi^4 \,d\varpi \,dl \,db
\end{equation}

We take into account the selection on the parallax relative uncertainty of 10\%, the $\vert Z \vert < 50$~pc constraint and the distance borders of the $E(B-V)<0.05$ contours. 
The parallax uncertainties are assumed to depend on the magnitude and we do not take into account the second order dependency on the colour nor on sky position. Consequently,

\begin{equation}
\begin{split}
  P(\Shrd | i,\varpi,l,b) &= P(\tilde{\varpi}/\sigma_\varpi>10 | M_G,\varpi) \\
  &\quad P(E(B-V)<0.05 | \varpi,l,b) \\
  &\quad P(\vert Z \vert < 50 | \varpi,l,b) \\
  &\quad P(G<20, \RP<19, \BP<18 | \varpi, i) 
\end{split}
\end{equation}

We model $\sigma_\varpi$ as a function of $G=M_G -5 - 5 \log(\varpi)$ by fitting a random sample representative of the Gaia data : $\sigma_\varpi = 0.023 + \exp(0.828 G-16.9)$ for $G$>13 and  $\sigma_\varpi =0.04$ for $G$<13. 

\begin{equation}
P(\tilde{\varpi}/\sigma_\varpi>10 | M_G,\varpi) = 1 - P(\tilde{\varpi} < 10 \sigma_\varpi | \varpi, \sigma_\varpi(G)) 
\end{equation}
the last term being determined by the cumulative distribution function of the Gaussian centred on $\varpi$ with dispersion $\sigma_\varpi$.

The $E(B-V)<0.05$ constraint corresponds to $d<d_{\rm max}(l,b)$, so 
\begin{equation}
 P(E(B-V)<0.05 | \varpi,l,b) = 1 - P( \tilde{\varpi}<1/d_{\rm max} | \varpi,d_{\rm max}(l,b))
\end{equation}

The $\vert Z \vert < 50$~pc constraint corresponds to $\varpi > \vert \sin b \vert / 0.05$.
\begin{equation}
 P(\vert Z \vert < 50 | \varpi,l,b) = 1 - P (\tilde{\varpi} < \frac{\vert \sin b \vert}{0.05}~|~\varpi, b)
\end{equation}

The maximum distance probed during the integration is set by the extinction constraint which corresponds to 1~kpc.

The photometric bands used in this study, $M_X$ = $M_{BP}$, $M_{RP}$, $M_J$, $M_H$, $M_K$ are added to the ($M_G$,$\bpmrp$) HRD using colour-colour relations $M_G-M_X$ as a function of $\bpmrp$. We chose a 7 order polynomial model for those relations and ensured that we did not extrapolate them. The calibrations are done using stars with extinction lower than $E(B-V)<0.01$~mag, photometric errors less than 2\% in the $G$ band and 5\% in $\BP$ and $\RP$, applying the photometric excess flux filter  \citep{Evans18}, $G>6$ to avoid saturation, $\BP<18$ to avoid background subtraction issues \citep{Evans18}, 2MASS photometric quality ``AAA'' and photometric errors in $J$,$H$ and $K$ smaller than 0.05~mag. As there are not enough very red stars with our strict criteria, we increased the extinction criterion to $E(B-V)<0.015$~mag for $\bpmrp>4$. 
We apply three different calibrations for (i) the red giants with $M_G<4.5$~mag and $M_G<-5.5+9(\bpmrp)$ (ii) the white dwarfs with $M_G<10+2.6(\bpmrp)$ and (iii) the dwarfs in between. To ensure the continuity between those calibrations, the red giant calibration has been derived with all stars with $M_G<2.5$, and the white dwarf relations have been derived using both the dwarf and the white dwarfs sample. As the different colour-colour relations are correlated, we fitted them simultaneously\footnote{using the R package systemfit} within each population. 

\begin{figure}[t]
\centering
\includegraphics[width=8cm]{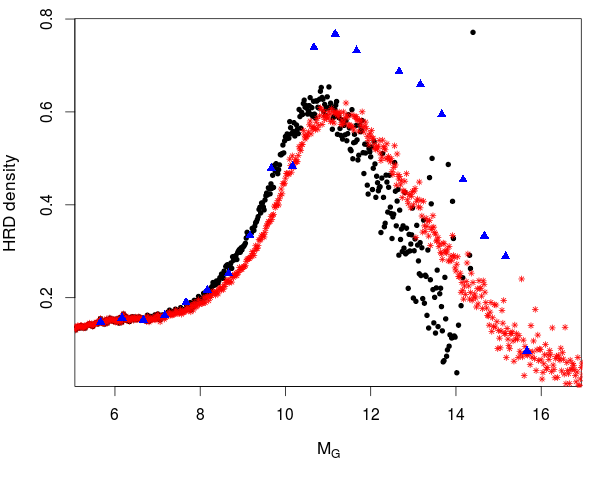}
\caption{Relative stellar densities along the HRD $P(M_G)$ as derived from Gaia data only ($P(M_G,\BP-\RP)$, black dots), Gaia and 2MASS ($P(M_G,J-K)$, red stars), Padova isochrones (blue triangles). The different densities are normalized at $M_G=6.5$.}
\label{fig:HRDproba}
\end{figure}

For a better modelling of the bottom of the HRD, quite important for the pollution of nearby red dwarfs in our CMDs, we applied the same procedure on a Gaia-2MASS HRD, $P(M_G, J-K)$, without constraint on $\BP$ nor $\RP$, using $J-K$ as the reference colour and selecting stars with $J<15.8$ and $K<14.3$ (which corresponds to the $>99\%$ completeness of the 2MASS catalogue \citep{Skrutskie06}). For the colour-colour relations of the faintest red dwarfs, we had to use the Phoenix relations \citep{Baraffe15} to derive the $\BP$ photometry for $J-K$>1.5. We merged the results of both HRDs at $M_G=6.5$. Figure~\ref{fig:HRDproba} shows the difference between both HRD densities for the bottom of the main sequence: at $M_G<11$ the higher resolution of Gaia leads to more intrinsically faint stars being observed than 2MASS while for fainter absolute magnitudes the $\BP$ quality criteria leads to too strong incompleteness in the Gaia data to be properly modelled. To confirm our interpretation of the differences we show on the same plot the HRD densities obtained with the theoretical HRD described in Sect.~\ref{ShrdIsoc}. However the exact density of low mass dwarfs has no implication on this work which concentrate on more distant stars, they only need to be present with the correct colour-colour relations to avoid them ending as strong outliers. 

\begin{figure}[t]
\centering
\includegraphics[width=8cm]{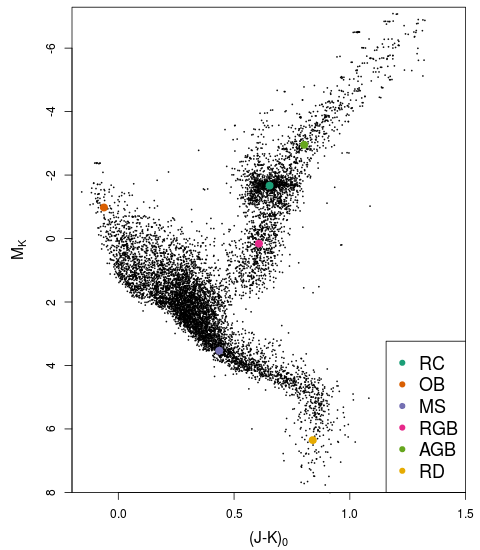}
\caption{Gaia DR2 empirical HR diagram. We overlay here some stars of different evolutionary stages for which we study their individual $P(O|\A,D)$ in appendix~\ref{Sannex}.
}
\label{fig:indivHR}
\end{figure}

\subsection{\label{ShrdIsoc}Theoretical H-R diagram prior}

To build a theoretical HRD, we use the PARSEC isochrones \citep{Bressan12} 
with a step of 0.1 Gyr in Age between [0.1, 13.4] and a step of 0.05 dex in [M/H] between [$-2.15$, 0.5]. Each isochrone point $i$, corresponding to a metallicity [M/H]$_i$, age $\tau_i$ and mass $\mathcal{M}_i$, has a weight associated to it $P(i)$ according to the IMF $P(\mathcal{M})$, an age distribution $P(\tau)$ and an age-metallicity relation (AMR) $P(\MH|\tau)$.
\begin{equation}
P(i) = P(\MH,\tau,\mathcal{M}) = P(\mathcal{M})\,P(\tau)\,P(\MH|\tau)  
\end{equation}
We use the \cite{Chabrier01} log-normal IMF (integrated over the mass interval between isochrone points), the \cite{RochaPinto00a} AMR and a constant SFR. We are observing here in the Galactic plane, so we can easily compute a rough correction to the relative number of stars of each age due to the different scale height $H_z$ of the populations as a function of age. Indeed if we assume that the density distribution at each age $\tau$ can be modelled under the assumption of an isothermal disc, the solution of the Jeans equation is then a sech$^2$ profile: 
\begin{equation}
\label{eq:rho}
\rho(z) = \rho_0\ \mathrm{sech}^2 ( \frac{z}{2 H_z})
\end{equation}
If we integrate over $z$ and assume that all the populations have the same radial density profile, we have 
\begin{equation}
\label{eq:Sigma}
 \Sigma = \int_z \rho(z)\,dz = 4\,\rho_0\,H_z
\end{equation}
and therefore a constant SFR corresponds to a local density 
\begin{equation}
P(\tau) \propto \rho_0  \propto 1 / H_z
\end{equation}
We are using here the default Hz of Trilegal 1.7 \citep{Girardi05}:  $H_z = 0.095 (1+\tau/5.55)^{1.6666}$. 

\begin{figure}[t]
\centering
\includegraphics[width=8cm]{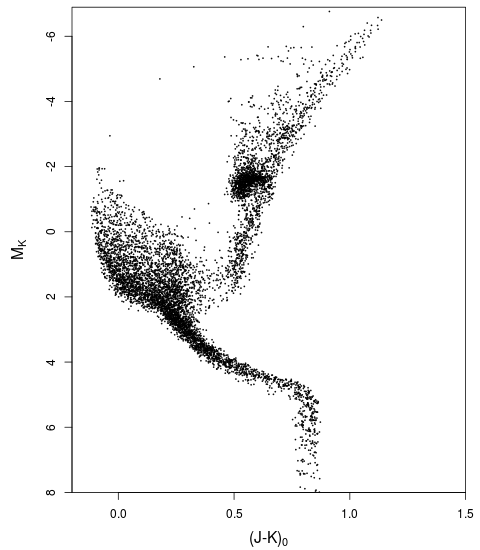}
\caption{Alternative HR diagram to Fig.~\ref{fig:indivHR} based on Padova isochrones.}
\label{fig:indivHRisocs}
\end{figure}

The resulting prior HRD is shown in Fig.~\ref{fig:indivHRisocs}.

\section{\label{sec:tests}Tests and results} 

We made extensive tests on our algorithm, first using simulations, then on real 2MASS and UKIDSS data combined with Gaia DR2. 
To check our results on real fields, we looked at a few fields where we knew what to expect, as the ones described below, and at several ones presenting different issues (high crowding, low stellar density towards the anti-centre, convergence issues...). For those, we checked how well our derived $\A(D)$ function permit to recover the red clump track. We also checked that both 2MASS and UKIDSS provided consistent results within the uncertainties. 

For 2MASS we selected stars with good photometric quality flags (A,B,C,or D). 
Following the prescription of \cite{Lucas08}, we correct the errors provided with the UKIDSS catalogue by the following:
\begin{equation}
\sigma_\mathrm{cor} = \sqrt{(1.2 \sigma)^2 + 0.02^2} 
\end{equation}
and we selected only stars with a photometric error lower than 0.1~mag. 
For the cross-match with Gaia DR2, we used the cross-match with 2MASS provided within Gaia DR2 \citep{Marrese19} and a simple cross-match within a radius of 0.15$\arcsec$ for UKIDSS. 
We applied the same Gaia photometric and astrometric filters as detailed in \cite{Paper2}: phot\_bp\_rp\_excess\_factor $> 1.3 +0.06$ (bp\_rp)$^2$, $G_{BP}>18$, astrometric\_chi2\_al/(astrometric\_n\_good\_obs\_al-5)$<$1.44 max(1,exp(-0.4*(G-19.5))), $\varpi + 3 \sigma_\varpi<0$, we take into account the 3 mmag/mag drift of the $G$ band, we add quadratically 10 mmag to the photometric uncertainties to take into account the systematics and we correct the parallax from the -0.03~mas zero point. 

\subsection{\label{Ssimu}Simulation}

\begin{figure}[t]
\centering
\includegraphics[width=\columnwidth]{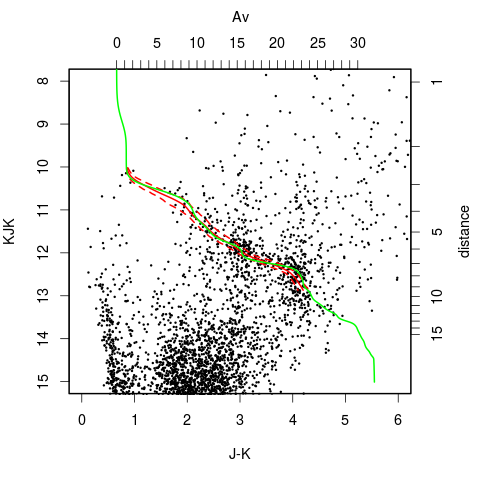}
\caption{CMD of the UKIDSS mock catalogue built from the Fux model stellar and gas particles distributions towards $l=10\degr$. The magnitude independent of extinction $K_{JK}$ is used. The corresponding distance and extinction for a Red Clump star on this diagram are indicated on the right and top axis.}
\label{fig:simuCMD}
\end{figure}

We tested our procedure on a simulation, as illustrated in Fig.~\ref{fig:resmat1outliers} and Fig.~\ref{fig:resmat2final}, corresponding to either 2MASS or UKIDSS observations towards $l=10\degr$. We simulated stars with intrinsic stellar properties randomly taken from the Hess diagram described in Sect.~\ref{ShrdGaia} and placed them along the line of sight following the \cite{Fux99} model stellar distance distribution and the cone effect. 
The extinction density is assumed to be proportional to the \cite{Fux99} model gas density in this direction and the proportion factor is simply derived assuming an integrated extinction along the line of sight of $\A^\infty$=32~mag. An intrinsic dispersion in the extinction is added using a log-normal distribution with $\sigma_A$=0.05. 2MASS, UKIDSS and Gaia photometric and parallax errors are added assuming a simple increase of the parallax errors with magnitude following a fit of Eq.~\ref{eq:sigm} on catalogue data. The completeness is then simulated following Eq.~\ref{eq:compl} with $\alpha=10$, $\beta=1$, $m_K^{*}=14.1$, $m_H^{*}=14.8$, $m_J^{*}=16.6$ for 2MASS and $\alpha=\beta=2$ and $m_K^{*}=19$, $m_H^{*}=19.5$, $m_J^{*}=22$ for UKIDSS. Gaia $G$ photometry and parallaxes are kept only if it satisfy the same completeness model as Eq.~\ref{eq:compl} with $m_G^{*}=20.7$ and $\BP$ and $\RP$ photometry with $m_{\BP}^{*}=20.9$ and $m_{\RP}^{*}=19.5$ (the exact values are not important as they do not enter the catalogue completeness model, but just allows us to take into account that Gaia information is not present for the faintest / reddest stars). 
We build this way mock catalogues of about 4\,000 stars satisfying our photometric criteria. 10\% of the UKIDSS mock catalogue has Gaia parallax information compared to 50\% for the 2MASS one. The UKIDSS mock catalogue is represented in Fig.~\ref{fig:simuCMD} using the magnitude independent of extinction $K_{JK}$ \citep[e.g. ][]{Babusiaux05}: 
\begin{equation}
K_{JK} = K - \frac{k_K}{k_J-k_K} (J-K) 
\end{equation}
with $k_J$ and $k_K$ the extinction coefficients in the $J$ and $K$ bands respectively.

\begin{figure}[t]
\centering
\includegraphics[width=\columnwidth]{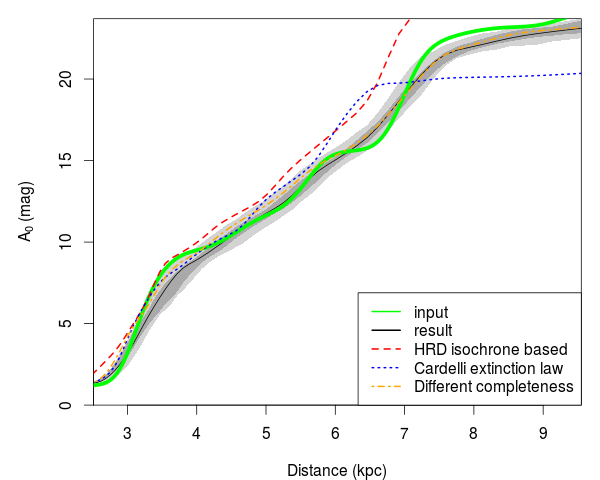}
\includegraphics[width=\columnwidth]{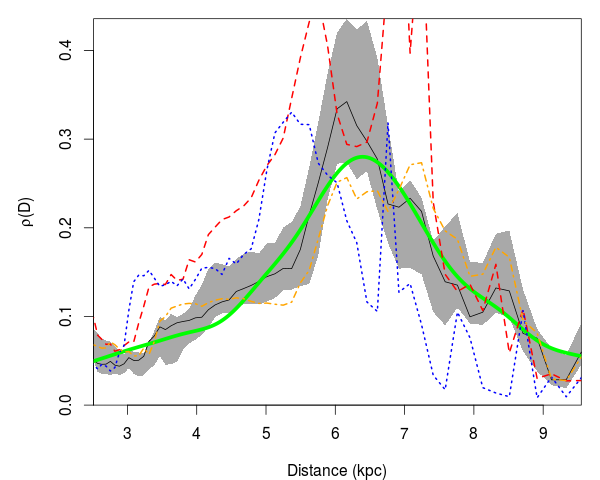}
\caption{Results of the deconvolution of the simulation for $\A(D)$ (top) and $\rho(D)$
(bottom) within the $\Dmin$/$\Dmax$ distance range. The dark line is the deconvolution result. The light grey area in the top panel
corresponds to the 1-$\sigma$ confidence interval of $\A(D)$ derived from the full $P(\A|D)$
(Fig.~\ref{fig:resmat2final}) while the darker grey area in both panels shows the 1-$\sigma$
confidence interval derived from the bootstrap. The green line is the input relation used in
the simulation. Red dashed line: isochrones HRD. Blue dotted line: \cite{Cardelli89}
extinction law. Orange dot-dashed line: assuming completeness parameters $\alpha=10$ and $\beta=1$.}
\label{fig:priorsinfluence}
\end{figure}

Figure~\ref{fig:priorsinfluence} shows in dark and grey the final results of the deconvolution on the UKIDSS mock catalogue. We see that the bootstrap confidence interval is smaller than the one derived directly from the full $P(\A|D)$ in Fig.~\ref{fig:resmat2final} which also takes into account the intrinsic dispersion of the extinction as well as the deconvolution artefacts and is therefore the confidence interval to be used. The residuals are within the 1-$\sigma$ confidence interval (with a dispersion of 0.55~mag) but are correlated by the fact that we impose a continuous increasing fit and the deconvolution artefacts.  
For the density the result is within the 2-$\sigma$ bootstrap confidence interval. Here again the residuals are correlated and correspond to an error of about 20\% on the density estimation.

We tested the influence of our choices of a number of parameters on the simulation. 

To test the HRD prior's influence, we processed our default simulation, done with the Gaia empirical HRD, using the isochrone-based HRD described in Sec.~\ref{ShrdIsoc}. We see in Fig.~\ref{fig:priorsinfluence} that the $\A(D)$ relation is reasonably well recovered although slightly shifted. The bar overdensity is still visible in the $\rho(D)$ distribution but is noisier. 

We tested the influence of the extinction law adopted by processing our default simulation, constructed with the \cite{FitzpatrickMassa07} extinction law, assuming in FEDReD the \cite{Cardelli89} extinction law.  We see in Fig.~\ref{fig:priorsinfluence} that the results are quite similar to the HRD change. 

Concerning our completeness model, FEDReD estimates magnitude limits slightly different from the input ones through the estimation using the maximum of the observed magnitude distribution, but they are still within 0.4~mag of the input ones. We checked that providing the exact input completeness values did not change sensibly the results. We also tested changing the $\alpha$ and $\beta$ parameters to 10 and 1 respectively (e.g. the 2MASS sharper values) for the UKIDSS simulation. Fig.~\ref{fig:priorsinfluence} shows that this affect, as expected, only $\rho(D)$ and not $\A(D)$.

\subsection{\label{SNGC}Field NGC 4815}

\begin{figure}[!ht]
\includegraphics[width=\columnwidth]{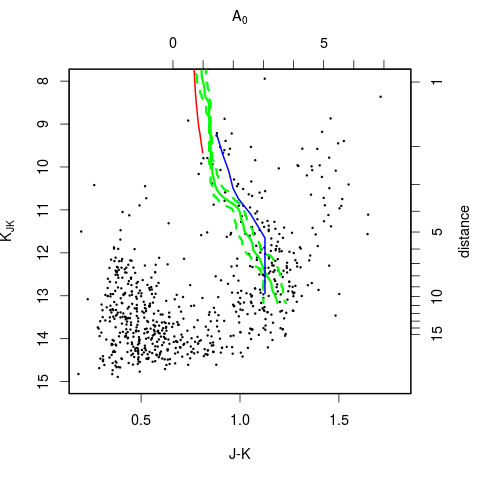}
\includegraphics[width=8cm]{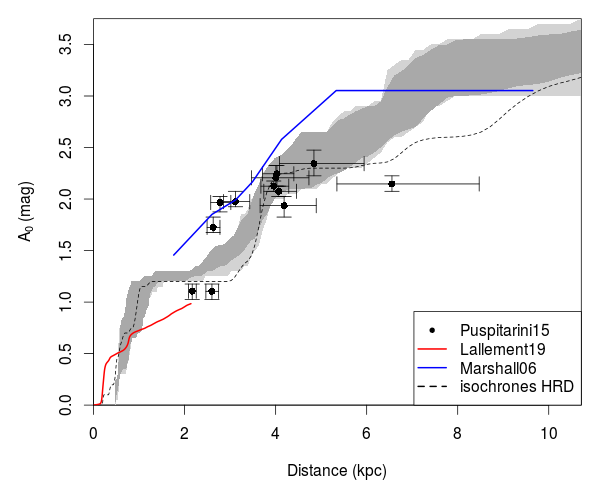}
\includegraphics[width=8cm]{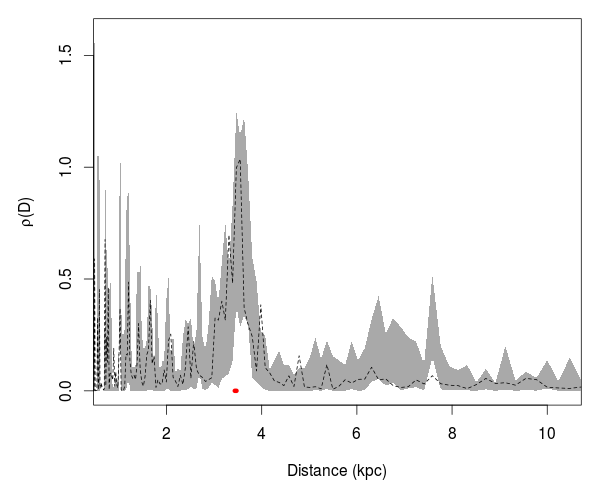}
\caption{Field of NGC~4815.  
Top: 2MASS CMD. In green the Red Clump track corresponding to our results with its 1-$\sigma$ confidence interval, in blue the \cite{Marshall06} results and in red the \cite{Lallement19} ones. 
Middle: Extinction, 1-$\sigma$ confidence interval in grey (see Fig.~\ref{fig:priorsinfluence}). Dotted line: FEDReD result using the isochrone HRD. Black points: \cite{Puspitarini15} updated with the Gaia DR2 distances, in red for members according to \citep{Friel14}. 
Bottom: stellar density. The cluster distance \citep{Cantat18} is indicated in red.
}
\label{fig:NGC4815}
\end{figure}

We looked at the FEDReD capabilities in the field around NGC 4815 studied in extinction with the Gaia-ESO Survey UVES observations by \cite{Puspitarini15}. We used 711 2MASS stars located in an area of 0.1$\degr$x0.1$\degr$ around $l=306.6\degr$, $b=-2.1\degr$, 87\% of those stars having Gaia parallaxes. This field is complex for FEDReD as it suffers from differential extinction requesting a very small field of view and therefore a small number of stars, and has the presence of a cluster which will differ from the mix of age and metallicities of our empirical HRD. To compare our results in Fig.~\ref{fig:NGC4815} with the ones of \cite{Puspitarini15}, we updated the distances of the latter with the Gaia DR2 distances using the inverse of the parallax. We also compare our results with the maps of \cite{Marshall06} and \cite{Lallement19}, our results being in between both maps with a better agreement with \cite{Puspitarini15}. This field is indicated as having a convergence issue in \cite{Green19}. 
The updated Gaia DR2 distances confirm that the 2 stars with lower extinction are foreground stars, as suspected by \cite{Puspitarini15}. We do not recover the same shape at small distances as \cite{Lallement19}, which can be due either to a too relaxed definition of $\Dmin$ from our side, considering the very few stars present in our small field of view to drive the solution, or to the too big resolution of the \cite{Lallement19} map for this specific area.
We confirm that a dust cloud is present at the cluster distance. We also confirm that the extinction continues to increase beyond the cluster, in phase with the higher velocity ISM structures seen in the HI data and not detected in the stars studied by \cite{Puspitarini15}. The extinction is likely to continue to increase beyond our distance limit as we do not reach the total extinction of 4.4~mag indicated by the map of \cite{Schlegel98}. Concerning the stellar density, we recover the overdensity linked to the presence of the cluster which we estimate to be at 3.5$\pm0.1$~kpc, which is consistent with the results of \cite{Cantat18}. Using the isochrone HRD instead of the empirical Gaia one leads to consistent results within the uncertainties.

\subsection{\label{S9P}Field 9P}

We looked at the capability of FEDReD to detect the bar signature using the field $l=9.55\degr,b=-0.09\degr$
studied in detail in \cite{Babusiaux05} with CIRSI near-infrared photometry and in \cite{Babusiaux14} with GIRAFFE spectroscopy. We took an area of 0.16$\degr\times0.16\degr$ leading to about 10\,000 UKIDSS stars with a photometric uncertainty lower than 0.1~mag in $J$, $H$ and $K$, e.g. up to $J=19$, $H=17.9$ and $K=16.8$~mag.
To improve the convergence at small distances, we completed UKIDSS with 2MASS photometry and we replaced the UKIDSS photometry by the 2MASS one for stars brighter than $J=13.25$, $H=12.75$, $K=12.0$, following \cite{Lucas08}.
For this we derived and applied colour-colour calibrations on well behaved stars of both surveys following \cite{Hodgkin09}. We used the isochrones generated with the UKIDSS filters and we transformed the empirical HRD from 2MASS to the UKIDSS photometric system using the transformations of \cite{Hodgkin09}. We used the same extinction coefficients as previously (i.e. \cite{Lallement19}).
In this field, the over-density in stellar counts due to the Galactic bar occurs brighter than the completeness limit. We checked through simulations that our default way to estimate the completeness parameters presented in Sect.~\ref{Scompl} is indeed adapted to this field.

\begin{figure}[!ht]
\centering
\includegraphics[width=\columnwidth]{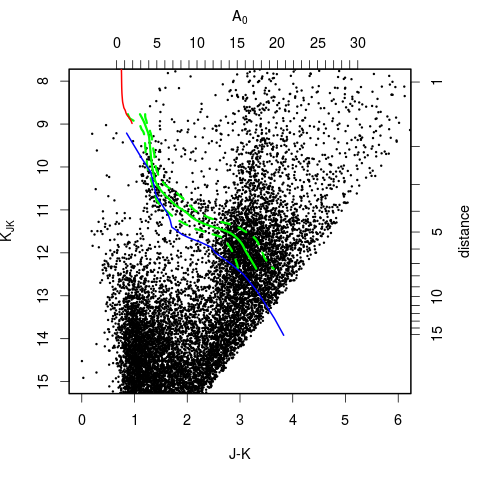}
\includegraphics[width=8cm]{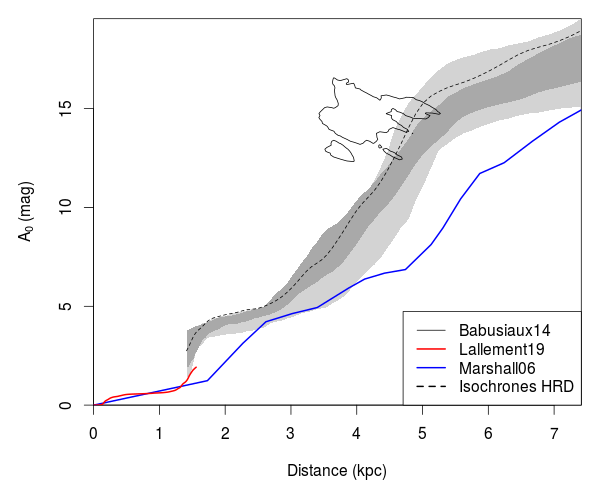}
\includegraphics[width=8cm]{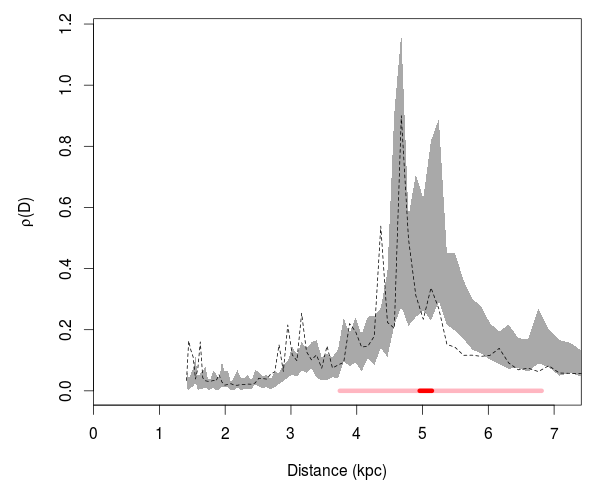}
\caption{Field $l=9.6\degr, b=0\degr$.
Top: UKIDSS CMD. In green the Red Clump track corresponding to our results with its 1-$\sigma$ confidence interval, in blue the \cite{Marshall06} results, in green the \cite{Lallement19} ones.
Middle: Extinction, 1-$\sigma$ confidence interval in grey (see Fig.~\ref{fig:priorsinfluence}). Dotted line: FEDReD result using the isochrone HRD. Thin back line: isocontours of the spectroscopic sample results of \cite{Babusiaux14}. 
Bottom: stellar density. The bar distance determined by \cite{Babusiaux05} is indicated with a red line and the distance spread in light red.}
\label{fig:l9.6}
\end{figure}

The results are presented in Fig.~\ref{fig:l9.6}. Both the empirical HRD and the synthetic one give consistent results within the uncertainties. Our results are barely overlapping in distance with the ones of \cite{Lallement19} but consistent within the uncertainties. We find a higher extinction than \cite{Marshall06}, more in agreement with the results of \cite{Babusiaux14}. The red clump track, clearly visible in the CMD, is well recovered.
We confirm the increase in extinction in the disk up to the bar location seen in \cite{Babusiaux14} and see the decrease of the extinction material afterwards. We see the location of the bar-driven overdensity at 4.9$\pm0.2$~kpc, which is consistent with
the value of \cite{Babusiaux05}. We also confirm the bar large distance spread. 
This large dispersion could be due to us seeing both the disk end and the bar, too close to be separated. The main increase in extinction seems to be slightly in front of the density peak, in agreement with what would be expected if we are seeing here the bar close to reaching the disk. The extension of the method to other longitudes to constrain the bar/disk interface will be presented in a forthcoming work.

\section{\label{sec:conclu}Conclusion}

We presented here a Bayesian deconvolution method, FEDReD, allowing us to derive at the same time the extinction distribution and stellar density maps taking into account the incompleteness of the surveys. 
We showed the performances of the algorithm on simulated data and on two test fields, one using 2MASS data centred around NGC~4815 and another using UKIDSS data towards the galactic bar at $l=10\degr$.  
The first full application of the method to construct an extinction map of the Galactic disk using 2MASS and Gaia DR2 is presented in \cite{Paper2}. Applications to UKIDSS and VVV data are underway. 

FEDReD is quite robust to differential extinction for its extinction derivation part, since it converges towards the median extinction behaviour. It is however important to select an homogeneous extinction behaviour to recover correctly the density distribution. We have seen towards NGC~4815 that it can work with a rather limited number of stars. 
Using the Gaia DR2 empirical HRD provides an accurate description of the local HRD, which we have shown to work well towards different parts of the disc. Still variations of the HRD within the disk (metallicity gradient, changing ratio thin/thick disk) can be preferred and implemented easily within FEDReD using the isochrone module.  
FEDReD has been designed to be flexible in its observable inputs so that any other knowledge for some stars of the field of view can be implemented such as spectroscopic and asteroseismology data.

\begin{acknowledgements}
We thank the referee for useful comments that helped to improve the paper.
This work has made use of data from the European
Space Agency (ESA) mission Gaia (https://www.cosmos.esa.int/gaia), processed by the Gaia Data Processing and Analysis Consortium (DPAC,
https://www.cosmos.esa.int/web/gaia/dpac/consortium). Funding for the DPAC
has been provided by national institutions, in particular the institutions participating in the Gaia Multilateral Agreement. 
This work makes use of data 
products from the 2MASS, which is a joint project of the University of Massachusetts and the Infrared Processing and Analysis Center/California Institute
of Technology, funded by the National Aeronautics and Space Administration
and the National Science Foundation.
This work is based in part on data obtained as part of the UKIRT Infrared Deep Sky Survey.
\end{acknowledgements}


\bibliographystyle{aa} 
\bibliography{chloe}

\begin{thebibliography}{57}
\expandafter\ifx\csname natexlab\endcsname\relax\def\natexlab#1{#1}\fi

\bibitem[{{Anders} {et~al.}(2019){Anders}, {Khalatyan}, {Chiappini}, {Queiroz},
  {Santiago}, {Jordi}, {Girardi}, {Brown}, {Matijevi{\v{c}}}, {Monari},
  {Cantat-Gaudin}, {Weiler}, {Khan}, {Miglio}, {Carrillo}, {Romero-G{\'o}mez},
  {Minchev}, {de Jong}, {Antoja}, {Ramos}, {Steinmetz}, \& {Enke}}]{Anders19}
{Anders}, F., {Khalatyan}, A., {Chiappini}, C., {et~al.} 2019, \aap, 628, A94

\bibitem[{{Arenou} {et~al.}(1992){Arenou}, {Grenon}, \& {Gomez}}]{Arenou92}
{Arenou}, F., {Grenon}, M., \& {Gomez}, A. 1992, \aap, 258, 104

\bibitem[{{Aringer} {et~al.}(2016){Aringer}, {Girardi}, {Nowotny}, {Marigo}, \&
  {Bressan}}]{Aringer16}
{Aringer}, B., {Girardi}, L., {Nowotny}, W., {Marigo}, P., \& {Bressan}, A.
  2016, \mnras, 457, 3611

\bibitem[{{Babusiaux} \& {Gilmore}(2005)}]{Babusiaux05}
{Babusiaux}, C. \& {Gilmore}, G. 2005, \mnras, 358, 1309

\bibitem[{{Babusiaux} {et~al.}(2014){Babusiaux}, {Katz}, {Hill}, {Royer},
  {G{\'o}mez}, {Arenou}, {Combes}, {Di Matteo}, {Gilmore}, {Haywood}, {Robin},
  {Rodriguez-Fernandez}, {Sartoretti}, \& {Schultheis}}]{Babusiaux14}
{Babusiaux}, C., {Katz}, D., {Hill}, V., {et~al.} 2014, \aap, 563, A15

\bibitem[{{Bailer-Jones}(2011)}]{CBJ11}
{Bailer-Jones}, C.~A.~L. 2011, \mnras, 411, 435

\bibitem[{{Baraffe} {et~al.}(2015){Baraffe}, {Homeier}, {Allard}, \&
  {Chabrier}}]{Baraffe15}
{Baraffe}, I., {Homeier}, D., {Allard}, F., \& {Chabrier}, G. 2015, \aap, 577,
  A42

\bibitem[{{Berry} {et~al.}(2012){Berry}, {Ivezi{\'c}}, {Sesar}, {Juri{\'c}},
  {Schlafly}, {Bellovary}, {Finkbeiner}, {Vrbanec}, {Beers}, {Brooks},
  {Schneider}, {Gibson}, {Kimball}, {Jones}, {Yoachim}, {Krughoff}, {Connolly},
  {Loebman}, {Bond}, {Schlegel}, {Dalcanton}, {Yanny}, {Majewski}, {Knapp},
  {Gunn}, {Allyn Smith}, {Fukugita}, {Kent}, {Barentine}, {Krzesinski}, \&
  {Long}}]{Berry12}
{Berry}, M., {Ivezi{\'c}}, {\v{Z}}., {Sesar}, B., {et~al.} 2012, \apj, 757, 166

\bibitem[{{Bressan} {et~al.}(2012){Bressan}, {Marigo}, {Girardi}, {Salasnich},
  {Dal Cero}, {Rubele}, \& {Nanni}}]{Bressan12}
{Bressan}, A., {Marigo}, P., {Girardi}, L., {et~al.} 2012, \mnras, 427, 127

\bibitem[{{Cabrera-Lavers} {et~al.}(2008){Cabrera-Lavers},
  {Gonz{\'a}lez-Fern{\'a}ndez}, {Garz{\'o}n}, {Hammersley}, \&
  {L{\'o}pez-Corredoira}}]{CabreraLavers08}
{Cabrera-Lavers}, A., {Gonz{\'a}lez-Fern{\'a}ndez}, C., {Garz{\'o}n}, F.,
  {Hammersley}, P.~L., \& {L{\'o}pez-Corredoira}, M. 2008, \aap, 491, 781

\bibitem[{{Cantat-Gaudin} {et~al.}(2018){Cantat-Gaudin}, {Jordi}, {Vallenari},
  {Bragaglia}, {Balaguer-N{\'u}{\~n}ez}, {Soubiran}, {Bossini}, {Moitinho},
  {Castro-Ginard}, {Krone-Martins}, {Casamiquela}, {Sordo}, \&
  {Carrera}}]{Cantat18}
{Cantat-Gaudin}, T., {Jordi}, C., {Vallenari}, A., {et~al.} 2018, \aap, 618,
  A93

\bibitem[{{Capitanio} {et~al.}(2017){Capitanio}, {Lallement}, {Vergely},
  {Elyajouri}, \& {Monreal-Ibero}}]{Capitanio17}
{Capitanio}, L., {Lallement}, R., {Vergely}, J.~L., {Elyajouri}, M., \&
  {Monreal-Ibero}, A. 2017, \aap, 606, A65

\bibitem[{{Cardelli} {et~al.}(1989){Cardelli}, {Clayton}, \&
  {Mathis}}]{Cardelli89}
{Cardelli}, J.~A., {Clayton}, G.~C., \& {Mathis}, J.~S. 1989, \apj, 345, 245

\bibitem[{{Chabrier}(2001)}]{Chabrier01}
{Chabrier}, G. 2001, \apj, 554, 1274

\bibitem[{{Chen} {et~al.}(2019){Chen}, {Huang}, {Yuan}, {Wang}, {Fan}, {Xiang},
  {Zhang}, {Tian}, \& {Liu}}]{Chen19}
{Chen}, B.-Q., {Huang}, Y., {Yuan}, H.-B., {et~al.} 2019, \mnras, 483, 4277

\bibitem[{{Chen} {et~al.}(2014){Chen}, {Liu}, {Yuan}, {Zhang}, {Schultheis},
  {Jiang}, {Huang}, {Xiang}, {Zhao}, {Yao}, \& {Lu}}]{Chen14}
{Chen}, B.-Q., {Liu}, X.-W., {Yuan}, H.-B., {et~al.} 2014, \mnras, 443, 1192

\bibitem[{{Chen} {et~al.}(2013){Chen}, {Schultheis}, {Jiang}, {Gonzalez},
  {Robin}, {Rejkuba}, \& {Minniti}}]{Chen13}
{Chen}, B.~Q., {Schultheis}, M., {Jiang}, B.~W., {et~al.} 2013, \aap, 550, A42

\bibitem[{{Danielski} {et~al.}(2018){Danielski}, {Babusiaux}, {Ruiz-Dern},
  {Sartoretti}, \& {Arenou}}]{Danielski18}
{Danielski}, C., {Babusiaux}, C., {Ruiz-Dern}, L., {Sartoretti}, P., \&
  {Arenou}, F. 2018, \aap, 614, A19

\bibitem[{{de Jong} {et~al.}(2008){de Jong}, {Rix}, {Martin}, {Zucker},
  {Dolphin}, {Bell}, {Belokurov}, \& {Evans}}]{deJong08}
{de Jong}, J.~T.~A., {Rix}, H.~W., {Martin}, N.~F., {et~al.} 2008, \aj, 135,
  1361

\bibitem[{{Drimmel} \& {Spergel}(2001)}]{Drimmel01}
{Drimmel}, R. \& {Spergel}, D.~N. 2001, \apj, 556, 181

\bibitem[{{Evans} {et~al.}(2018){Evans}, {Riello}, {De Angeli}, {Carrasco},
  {Montegriffo}, {Fabricius}, {Jordi}, {Palaversa}, {Diener}, {Busso},
  {Cacciari}, {van Leeuwen}, {Burgess}, {Davidson}, {Harrison}, {Hodgkin},
  {Pancino}, {Richards}, {Altavilla}, {Balaguer-N{\'u}{\~n}ez}, {Barstow},
  {Bellazzini}, {Brown}, {Castellani}, {Cocozza}, {De Luise}, {Delgado},
  {Ducourant}, {Galleti}, {Gilmore}, {Giuffrida}, {Holl}, {Kewley}, {Koposov},
  {Marinoni}, {Marrese}, {Osborne}, {Piersimoni}, {Portell}, {Pulone},
  {Ragaini}, {Sanna}, {Terrett}, {Walton}, {Wevers}, \&
  {Wyrzykowski}}]{Evans18}
{Evans}, D.~W., {Riello}, M., {De Angeli}, F., {et~al.} 2018, \aap, 616, A4

\bibitem[{{Fitzpatrick} \& {Massa}(2007)}]{FitzpatrickMassa07}
{Fitzpatrick}, E.~L. \& {Massa}, D. 2007, \apj, 663, 320

\bibitem[{{Friel} {et~al.}(2014){Friel}, {Donati}, {Bragaglia}, {Jacobson},
  {Magrini}, {Prisinzano}, {Rand ich}, {Tosi}, {Cantat-Gaudin}, {Vallenari},
  {Smiljanic}, {Carraro}, {Sordo}, {Maiorca}, {Tautvai{\v{s}}ien{\.{e}}},
  {Sestito}, {Zaggia}, {Jim{\'e}nez-Esteban}, {Gilmore}, {Jeffries}, {Alfaro},
  {Bensby}, {Koposov}, {Korn}, {Pancino}, {Recio-Blanco}, {Franciosini},
  {Hill}, {Jackson}, {de Laverny}, {Morbidelli}, {Sacco}, {Worley},
  {Hourihane}, {Costado}, {Jofr{\'e}}, \& {Lind}}]{Friel14}
{Friel}, E.~D., {Donati}, P., {Bragaglia}, A., {et~al.} 2014, \aap, 563, A117

\bibitem[{{Fux}(1999)}]{Fux99}
{Fux}, R. 1999, \aap, 345, 787

\bibitem[{{Gaia Collaboration} {et~al.}(2018{\natexlab{a}}){Gaia
  Collaboration}, {Babusiaux}, {van Leeuwen}, {Barstow}, {Jordi}, {Vallenari},
  {Bossini}, {Bressan}, {Cantat-Gaudin}, {van Leeuwen}, \& et~al.}]{DPACP-31}
{Gaia Collaboration}, {Babusiaux}, C., {van Leeuwen}, F., {et~al.}
  2018{\natexlab{a}}, \aap, 616, A10

\bibitem[{{Gaia Collaboration} {et~al.}(2018{\natexlab{b}}){Gaia
  Collaboration}, {Brown}, {Vallenari}, {Prusti}, {de Bruijne}, {Babusiaux},
  {Bailer-Jones}, {Biermann}, {Evans}, {Eyer}, {Jansen}, {Jordi}, {Klioner},
  {Lammers}, {Lindegren}, {Luri}, {Mignard}, {Panem}, {Pourbaix}, {Randich},
  {Sartoretti}, {Siddiqui}, {Soubiran}, {van Leeuwen}, {Walton}, {Arenou},
  {Bastian}, {Cropper}, {Drimmel}, {Katz}, {Lattanzi}, {Bakker}, {Cacciari},
  {Casta{\~n}eda}, {Chaoul}, {Cheek}, {De Angeli}, {Fabricius}, {Guerra},
  {Holl}, {Masana}, {Messineo}, {Mowlavi}, {Nienartowicz}, {Panuzzo},
  {Portell}, {Riello}, {Seabroke}, {Tanga}, {Th{\'e}venin}, {Gracia-Abril},
  {Comoretto}, {Garcia-Reinaldos}, {Teyssier}, {Altmann}, {Andrae}, {Audard},
  {Bellas-Velidis}, {Benson}, {Berthier}, {Blomme}, {Burgess}, {Busso},
  {Carry}, {Cellino}, {Clementini}, {Clotet}, {Creevey}, {Davidson}, {De
  Ridder}, {Delchambre}, {Dell'Oro}, {Ducourant},
  {Fern{\'a}ndez-Hern{\'a}ndez}, {Fouesneau}, {Fr{\'e}mat}, {Galluccio},
  {Garc{\'\i}a-Torres}, {Gonz{\'a}lez-N{\'u}{\~n}ez}, {Gonz{\'a}lez-Vidal},
  {Gosset}, {Guy}, {Halbwachs}, {Hambly}, {Harrison}, {Hern{\'a}ndez},
  {Hestroffer}, {Hodgkin}, {Hutton}, {Jasniewicz}, {Jean-Antoine-Piccolo},
  {Jordan}, {Korn}, {Krone-Martins}, {Lanzafame}, {Lebzelter}, {L{\"o}ffler},
  {Manteiga}, {Marrese}, {Mart{\'\i}n-Fleitas}, {Moitinho}, {Mora}, {Muinonen},
  {Osinde}, {Pancino}, {Pauwels}, {Petit}, {Recio-Blanco}, {Richards},
  {Rimoldini}, {Robin}, {Sarro}, {Siopis}, {Smith}, {Sozzetti}, {S{\"u}veges},
  {Torra}, {van Reeven}, {Abbas}, {Abreu Aramburu}, {Accart}, {Aerts},
  {Altavilla}, {{\'A}lvarez}, {Alvarez}, {Alves}, {Anderson}, {Andrei},
  {Anglada Varela}, {Antiche}, {Antoja}, {Arcay}, {Astraatmadja}, {Bach},
  {Baker}, {Balaguer-N{\'u}{\~n}ez}, {Balm}, {Barache}, {Barata}, {Barbato},
  {Barblan}, {Barklem}, {Barrado}, {Barros}, {Barstow}, {Bartholom{\'e}
  Mu{\~n}oz}, {Bassilana}, {Becciani}, {Bellazzini}, {Berihuete}, {Bertone},
  {Bianchi}, {Bienaym{\'e}}, {Blanco-Cuaresma}, {Boch}, {Boeche}, {Bombrun},
  {Borrachero}, {Bossini}, {Bouquillon}, {Bourda}, {Bragaglia}, {Bramante},
  {Breddels}, {Bressan}, {Brouillet}, {Br{\"u}semeister}, {Brugaletta},
  {Bucciarelli}, {Burlacu}, {Busonero}, {Butkevich}, {Buzzi}, {Caffau},
  {Cancelliere}, {Cannizzaro}, {Cantat-Gaudin}, {Carballo}, {Carlucci},
  {Carrasco}, {Casamiquela}, {Castellani}, {Castro-Ginard}, {Charlot},
  {Chemin}, {Chiavassa}, {Cocozza}, {Costigan}, {Cowell}, {Crifo}, {Crosta},
  {Crowley}, {Cuypers}, {Dafonte}, {Damerdji}, {Dapergolas}, {David}, {David},
  {de Laverny}, {De Luise}, {De March}, {de Martino}, {de Souza}, {de Torres},
  {Debosscher}, {del Pozo}, {Delbo}, {Delgado}, {Delgado}, {Di Matteo},
  {Diakite}, {Diener}, {Distefano}, {Dolding}, {Drazinos}, {Dur{\'a}n},
  {Edvardsson}, {Enke}, {Eriksson}, {Esquej}, {Eynard Bontemps}, {Fabre},
  {Fabrizio}, {Faigler}, {Falc{\~a}o}, {Farr{\`a}s Casas}, {Federici},
  {Fedorets}, {Fernique}, {Figueras}, {Filippi}, {Findeisen}, {Fonti},
  {Fraile}, {Fraser}, {Fr{\'e}zouls}, {Gai}, {Galleti}, {Garabato},
  {Garc{\'\i}a-Sedano}, {Garofalo}, {Garralda}, {Gavel}, {Gavras}, {Gerssen},
  {Geyer}, {Giacobbe}, {Gilmore}, {Girona}, {Giuffrida}, {Glass}, {Gomes},
  {Granvik}, {Gueguen}, {Guerrier}, {Guiraud}, {Guti{\'e}rrez-S{\'a}nchez},
  {Haigron}, {Hatzidimitriou}, {Hauser}, {Haywood}, {Heiter}, {Helmi}, {Heu},
  {Hilger}, {Hobbs}, {Hofmann}, {Holland}, {Huckle}, {Hypki}, {Icardi},
  {Jan{\ss}en}, {Jevardat de Fombelle}, {Jonker}, {Juh{\'a}sz}, {Julbe},
  {Karampelas}, {Kewley}, {Klar}, {Kochoska}, {Kohley}, {Kolenberg},
  {Kontizas}, {Kontizas}, {Koposov}, {Kordopatis}, {Kostrzewa-Rutkowska},
  {Koubsky}, {Lambert}, {Lanza}, {Lasne}, {Lavigne}, {Le Fustec}, {Le
  Poncin-Lafitte}, {Lebreton}, {Leccia}, {Leclerc}, {Lecoeur-Taibi},
  {Lenhardt}, {Leroux}, {Liao}, {Licata}, {Lindstr{\o}m}, {Lister}, {Livanou},
  {Lobel}, {L{\'o}pez}, {Managau}, {Mann}, {Mantelet}, {Marchal}, {Marchant},
  {Marconi}, {Marinoni}, {Marschalk{\'o}}, {Marshall}, {Martino}, {Marton},
  {Mary}, {Massari}, {Matijevi{\v{c}}}, {Mazeh}, {McMillan}, {Messina},
  {Michalik}, {Millar}, {Molina}, {Molinaro}, {Moln{\'a}r}, {Montegriffo},
  {Mor}, {Morbidelli}, {Morel}, {Morris}, {Mulone}, {Muraveva}, {Musella},
  {Nelemans}, {Nicastro}, {Noval}, {O'Mullane}, {Ord{\'e}novic},
  {Ord{\'o}{\~n}ez-Blanco}, {Osborne}, {Pagani}, {Pagano}, {Pailler},
  {Palacin}, {Palaversa}, {Panahi}, {Pawlak}, {Piersimoni}, {Pineau}, {Plachy},
  {Plum}, {Poggio}, {Poujoulet}, {Pr{\v{s}}a}, {Pulone}, {Racero}, {Ragaini},
  {Rambaux}, {Ramos-Lerate}, {Regibo}, {Reyl{\'e}}, {Riclet}, {Ripepi}, {Riva},
  {Rivard}, {Rixon}, {Roegiers}, {Roelens}, {Romero-G{\'o}mez}, {Rowell},
  {Royer}, {Ruiz-Dern}, {Sadowski}, {Sagrist{\`a} Sell{\'e}s}, {Sahlmann},
  {Salgado}, {Salguero}, {Sanna}, {Santana-Ros}, {Sarasso}, {Savietto},
  {Schultheis}, {Sciacca}, {Segol}, {Segovia}, {S{\'e}gransan}, {Shih},
  {Siltala}, {Silva}, {Smart}, {Smith}, {Solano}, {Solitro}, {Sordo}, {Soria
  Nieto}, {Souchay}, {Spagna}, {Spoto}, {Stampa}, {Steele},
  {Steidelm{\"u}ller}, {Stephenson}, {Stoev}, {Suess}, {Surdej}, {Szabados},
  {Szegedi-Elek}, {Tapiador}, {Taris}, {Tauran}, {Taylor}, {Teixeira},
  {Terrett}, {Teyssand ier}, {Thuillot}, {Titarenko}, {Torra Clotet}, {Turon},
  {Ulla}, {Utrilla}, {Uzzi}, {Vaillant}, {Valentini}, {Valette}, {van Elteren},
  {Van Hemelryck}, {van Leeuwen}, {Vaschetto}, {Vecchiato}, {Veljanoski},
  {Viala}, {Vicente}, {Vogt}, {von Essen}, {Voss}, {Votruba}, {Voutsinas},
  {Walmsley}, {Weiler}, {Wertz}, {Wevers}, {Wyrzykowski}, {Yoldas},
  {{\v{Z}}erjal}, {Ziaeepour}, {Zorec}, {Zschocke}, {Zucker}, {Zurbach}, \&
  {Zwitter}}]{GDR2paper}
{Gaia Collaboration}, {Brown}, A.~G.~A., {Vallenari}, A., {et~al.}
  2018{\natexlab{b}}, \aap, 616, A1

\bibitem[{{Gaia~Collaboration} {et~al.}(2016){Gaia~Collaboration}, {Prusti},
  {de Bruijne}, {Brown}, {Vallenari}, {Babusiaux}, {Bailer-Jones}, {Bastian},
  {Biermann}, {Evans}, \& et~al.}]{Prusti16}
{Gaia~Collaboration}, {Prusti}, T., {de Bruijne}, J.~H.~J., {et~al.} 2016,
  \aap, 595, A1

\bibitem[{{Girardi} {et~al.}(2005){Girardi}, {Groenewegen}, {Hatziminaoglou},
  \& {da Costa}}]{Girardi05}
{Girardi}, L., {Groenewegen}, M.~A.~T., {Hatziminaoglou}, E., \& {da Costa}, L.
  2005, \aap, 436, 895

\bibitem[{{Gontcharov}(2017)}]{Gontcharov17}
{Gontcharov}, G.~A. 2017, Astronomy Letters, 43, 472

\bibitem[{{Green} {et~al.}(2014){Green}, {Schlafly}, {Finkbeiner}, {Juri{\'c}},
  {Rix}, {Burgett}, {Chambers}, {Draper}, {Flewelling}, {Kudritzki}, {Magnier},
  {Martin}, {Metcalfe}, {Tonry}, {Wainscoat}, \& {Waters}}]{Green14}
{Green}, G.~M., {Schlafly}, E.~F., {Finkbeiner}, D.~P., {et~al.} 2014, \apj,
  783, 114

\bibitem[{{Green} {et~al.}(2019){Green}, {Schlafly}, {Zucker}, {Speagle}, \&
  {Finkbeiner}}]{Green19}
{Green}, G.~M., {Schlafly}, E.~F., {Zucker}, C., {Speagle}, J.~S., \&
  {Finkbeiner}, D.~P. 2019, arXiv e-prints, arXiv:1905.02734

\bibitem[{{Hanson} \& {Bailer-Jones}(2014)}]{Hanson14}
{Hanson}, R.~J. \& {Bailer-Jones}, C.~A.~L. 2014, \mnras, 438, 2938

\bibitem[{{Hodgkin} {et~al.}(2009){Hodgkin}, {Irwin}, {Hewett}, \&
  {Warren}}]{Hodgkin09}
{Hodgkin}, S.~T., {Irwin}, M.~J., {Hewett}, P.~C., \& {Warren}, S.~J. 2009,
  \mnras, 394, 675

\bibitem[{Hottier {et~al.}(2020)Hottier, C., \& F.}]{Paper2}
Hottier, C., C., B., \& F., A. 2020, in prep

\bibitem[{{Lallement} {et~al.}(2019){Lallement}, {Babusiaux}, {Vergely},
  {Katz}, {Arenou}, {Valette}, {Hottier}, \& {Capitanio}}]{Lallement19}
{Lallement}, R., {Babusiaux}, C., {Vergely}, J.~L., {et~al.} 2019, arXiv
  e-prints

\bibitem[{{L{\'o}pez-Corredoira} {et~al.}(2000){L{\'o}pez-Corredoira},
  {Hammersley}, {Garz{\'o}n}, {Simonneau}, \& {Mahoney}}]{LopezCorredoira00}
{L{\'o}pez-Corredoira}, M., {Hammersley}, P.~L., {Garz{\'o}n}, F., {Simonneau},
  E., \& {Mahoney}, T.~J. 2000, \mnras, 313, 392

\bibitem[{{Lucas} {et~al.}(2008){Lucas}, {Hoare}, {Longmore}, {Schr{\"o}der},
  {Davis}, {Adamson}, {Bandyopadhyay}, {de Grijs}, {Smith}, {Gosling},
  {Mitchison}, {G{\'a}sp{\'a}r}, {Coe}, {Tamura}, {Parker}, {Irwin}, {Hambly},
  {Bryant}, {Collins}, {Cross}, {Evans}, {Gonzalez-Solares}, {Hodgkin},
  {Lewis}, {Read}, {Riello}, {Sutorius}, {Lawrence}, {Drew}, {Dye}, \&
  {Thompson}}]{Lucas08}
{Lucas}, P.~W., {Hoare}, M.~G., {Longmore}, A., {et~al.} 2008, \mnras, 391, 136

\bibitem[{{Lucy}(1974)}]{Lucy74}
{Lucy}, L.~B. 1974, \aj, 79, 745

\bibitem[{{Marrese} {et~al.}(2019){Marrese}, {Marinoni}, {Fabrizio}, \&
  {Altavilla}}]{Marrese19}
{Marrese}, P.~M., {Marinoni}, S., {Fabrizio}, M., \& {Altavilla}, G. 2019,
  \aap, 621, A144

\bibitem[{{Marshall} {et~al.}(2006){Marshall}, {Robin}, {Reyl{\'e}},
  {Schultheis}, \& {Picaud}}]{Marshall06}
{Marshall}, D.~J., {Robin}, A.~C., {Reyl{\'e}}, C., {Schultheis}, M., \&
  {Picaud}, S. 2006, \aap, 453, 635

\bibitem[{Ng \& Maechler(2007)}]{cobs}
Ng, P. \& Maechler, M. 2007, Statistical Modelling, 7, 315

\bibitem[{{Nishiyama} {et~al.}(2005){Nishiyama}, {Nagata}, {Baba}, {Haba},
  {Kadowaki}, {Kato}, {Kurita}, {Nagashima}, {Nagayama}, {Murai}, {Nakajima},
  {Tamura}, {Nakaya}, {Sugitani}, {Naoi}, {Matsunaga}, {Tanab{\'e}},
  {Kusakabe}, \& {Sato}}]{Nishiyama05}
{Nishiyama}, S., {Nagata}, T., {Baba}, D., {et~al.} 2005, \apjl, 621, L105

\bibitem[{{Puspitarini} {et~al.}(2015){Puspitarini}, {Lallement}, {Babusiaux},
  {Chen}, {Bonifacio}, {Sbordone}, {Caffau}, {Duffau}, {Hill}, {Monreal-Ibero},
  {Royer}, {Arenou}, {Peralta}, {Drew}, {Bonito}, {Lopez-Santiago}, {Alfaro},
  {Bensby}, {Bragaglia}, {Flaccomio}, {Lanzafame}, {Pancino}, {Recio-Blanco},
  {Smiljanic}, {Costado}, {Lardo}, {de Laverny}, \& {Zwitter}}]{Puspitarini15}
{Puspitarini}, L., {Lallement}, R., {Babusiaux}, C., {et~al.} 2015, \aap, 573,
  A35

\bibitem[{{Reyl{\'e}} {et~al.}(2009){Reyl{\'e}}, {Marshall}, {Robin}, \&
  {Schultheis}}]{Reyle09}
{Reyl{\'e}}, C., {Marshall}, D.~J., {Robin}, A.~C., \& {Schultheis}, M. 2009,
  \aap, 495, 819

\bibitem[{{Rezaei Kh.} {et~al.}(2017){Rezaei Kh.}, {Bailer-Jones}, {Hanson}, \&
  {Fouesneau}}]{Rezaei17}
{Rezaei Kh.}, S., {Bailer-Jones}, C.~A.~L., {Hanson}, R.~J., \& {Fouesneau}, M.
  2017, \aap, 598, A125

\bibitem[{{Richardson}(1972)}]{Richardson72}
{Richardson}, W.~H. 1972, Journal of the Optical Society of America
  (1917-1983), 62, 55

\bibitem[{{Rocha-Pinto} {et~al.}(2000){Rocha-Pinto}, {Maciel}, {Scalo}, \&
  {Flynn}}]{RochaPinto00a}
{Rocha-Pinto}, H.~J., {Maciel}, W.~J., {Scalo}, J., \& {Flynn}, C. 2000, \aap,
  358, 850

\bibitem[{{Ruiz-Dern} {et~al.}(2018){Ruiz-Dern}, {Babusiaux}, {Arenou},
  {Turon}, \& {Lallement}}]{RuizDern18}
{Ruiz-Dern}, L., {Babusiaux}, C., {Arenou}, F., {Turon}, C., \& {Lallement}, R.
  2018, \aap, 609, A116

\bibitem[{{Sale}(2012)}]{Sale12}
{Sale}, S.~E. 2012, \mnras, 427, 2119

\bibitem[{{Schlegel} {et~al.}(1998){Schlegel}, {Finkbeiner}, \&
  {Davis}}]{Schlegel98}
{Schlegel}, D.~J., {Finkbeiner}, D.~P., \& {Davis}, M. 1998, \apj, 500, 525

\bibitem[{{Schultheis} {et~al.}(2014){Schultheis}, {Chen}, {Jiang}, {Gonzalez},
  {Enokiya}, {Fukui}, {Torii}, {Rejkuba}, \& {Minniti}}]{Schultheis14}
{Schultheis}, M., {Chen}, B.~Q., {Jiang}, B.~W., {et~al.} 2014, \aap, 566, A120

\bibitem[{{Skrutskie} {et~al.}(2006){Skrutskie}, {Cutri}, {Stiening},
  {Weinberg}, {Schneider}, {Carpenter}, {Beichman}, {Capps}, {Chester},
  {Elias}, {Huchra}, {Liebert}, {Lonsdale}, {Monet}, {Price}, {Seitzer},
  {Jarrett}, {Kirkpatrick}, {Gizis}, {Howard}, {Evans}, {Fowler}, {Fullmer},
  {Hurt}, {Light}, {Kopan}, {Marsh}, {McCallon}, {Tam}, {Van Dyk}, \&
  {Wheelock}}]{Skrutskie06}
{Skrutskie}, M.~F., {Cutri}, R.~M., {Stiening}, R., {et~al.} 2006, \aj, 131,
  1163

\bibitem[{{Stanek} {et~al.}(1994){Stanek}, {Mateo}, {Udalski}, {Szymanski},
  {Kaluzny}, \& {Kubiak}}]{Stanek94}
{Stanek}, K.~Z., {Mateo}, M., {Udalski}, A., {et~al.} 1994, \apjl, 429, L73

\bibitem[{{Surot} {et~al.}(2019){Surot}, {Valenti}, {Hidalgo}, {Zoccali},
  {Gonzalez}, {S{\"o}kmen}, {Minniti}, {Rejkuba}, \& {Lucas}}]{Surot19}
{Surot}, F., {Valenti}, E., {Hidalgo}, S.~L., {et~al.} 2019, arXiv e-prints,
  arXiv:1907.01972

\bibitem[{{Vergely} {et~al.}(2010){Vergely}, {Valette}, {Lallement}, \&
  {Raimond}}]{Vergely10}
{Vergely}, J.-L., {Valette}, B., {Lallement}, R., \& {Raimond}, S. 2010, \aap,
  518, A31

\bibitem[{{Wegg} \& {Gerhard}(2013)}]{Wegg13}
{Wegg}, C. \& {Gerhard}, O. 2013, \mnras, 435, 1874

\bibitem[{{Wegg} {et~al.}(2015){Wegg}, {Gerhard}, \& {Portail}}]{Wegg15}
{Wegg}, C., {Gerhard}, O., \& {Portail}, M. 2015, \mnras, 450, 4050

\end{thebibliography}

\twocolumn


\pagebreak

\onecolumn

\appendix

\section{\label{Sannex}Individual $P(\Os|\A,D)$}

\begin{figure*}[h]
\centering
\includegraphics[width=0.9\textwidth]{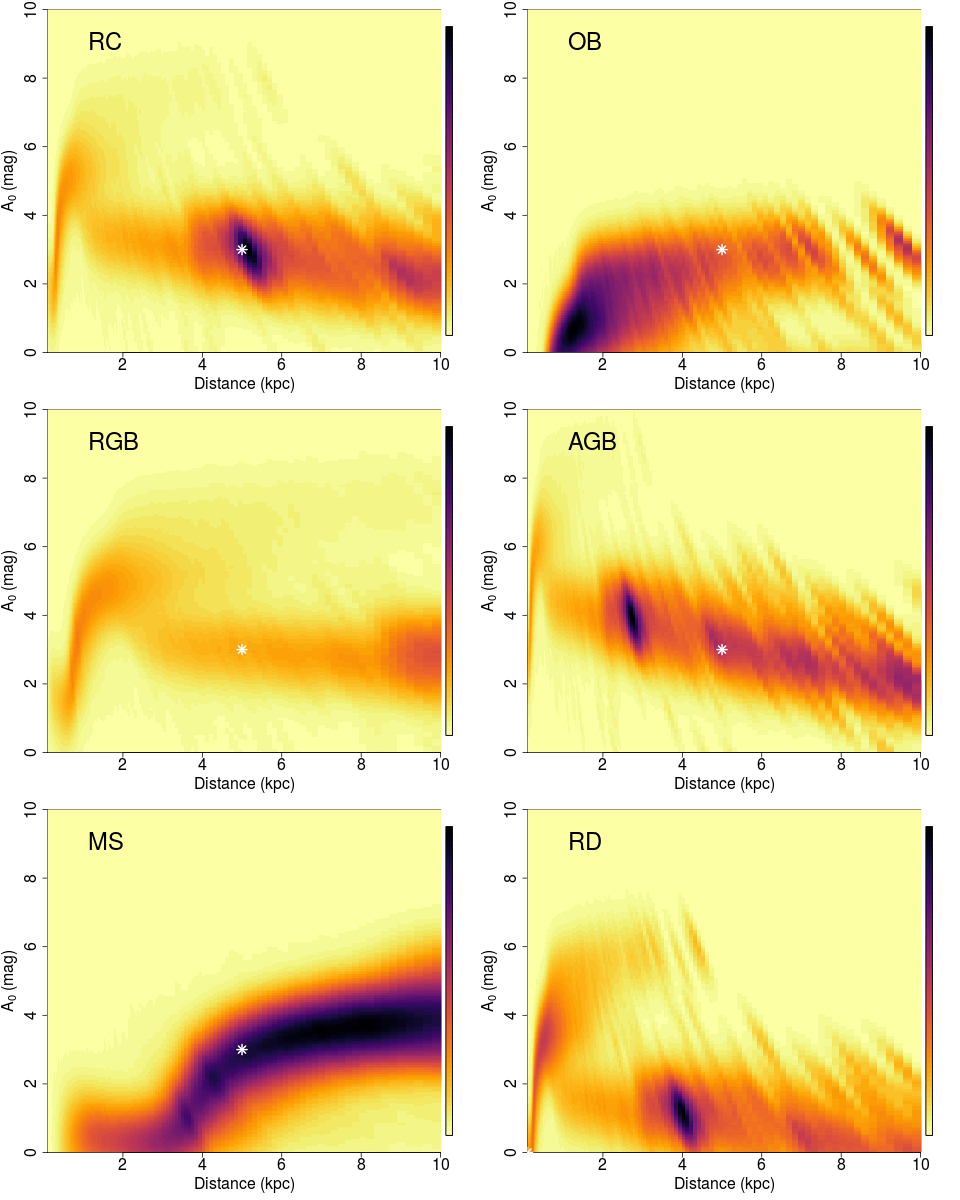}
\caption{$P(\Os|\A,D)P_0(D)$ for various stellar types indicated in Fig.~\ref{fig:indivHR} using as observables $J,H,K$ photometry only, displayed with a square-root colour scale. The prior on stellar density is chosen here to be uniform, e.g. containing only the cone effect: $P(D)\propto D^2$. 
All stars are located at 4~kpc with an extinction $\A=3$~mag, with the exception of the Red Dwarf, which is located at 0.1~kpc without extinction. The real position of the star is indicated by a white point. We see in this plot that the information is mostly carried by the Red Clump stars and that the Gaia parallax and/or photometry is needed to differentiate a red clump star from a red dwarf. }
\label{fig:indivPAD}
\end{figure*}

\end{document}